\documentclass[twocolumn]{aastex62}
\usepackage{graphicx}
\usepackage{natbib}
\usepackage{amsmath}
\usepackage{rotating}

\newcommand{\OIII}{\hbox{{\rm [O}\kern 0.1em{\sc iii}{\rm ]}}}

\newcommand{\UVJ}{\textit{UVJ}}
\newcommand{\UVc}{\textit{(U-V)}}
\newcommand{\VJc}{\textit{(V-J)}}

\newcommand{\logM}{$\log(M/{\rm M}_\odot)$}

\begin{document}

\title{An Extremely Massive Quiescent Galaxy at $z=3.493$: Evidence of Insufficiently Rapid Quenching Mechanisms in Theoretical Models\footnote{The spectroscopic data presented herein were obtained at the W. M. Keck Observatory, which is operated as a scientific partnership among the California Institute of Technology, the University of California and the National Aeronautics and Space Administration. The Observatory was made possible by the generous financial support of the W. M. Keck Foundation.}}

\correspondingauthor{Ben Forrest}
\email{benjamif@ucr.edu}

\author[0000-0001-6003-0541]{Ben Forrest}
	\affiliation{Department of Physics and Astronomy, University of California, Riverside, CA 92521, USA}
\author{Marianna Annunziatella}
	\affiliation{Physics Department, Tufts University, Medford, MA, USA}
\author[0000-0002-6572-7089]{Gillian Wilson}
	\affiliation{Department of Physics and Astronomy, University of California, Riverside, CA 92521, USA}
\author[0000-0001-9002-3502]{Danilo Marchesini}
	\affiliation{Physics Department, Tufts University, Medford, MA, USA}
\author[0000-0002-9330-9108]{Adam Muzzin}
	\affiliation{Department of Physics and Astronomy, York University, Toronto, Ontario, Canada}
\author[0000-0003-1371-6019]{M. C. Cooper}
	\affiliation{Center for Cosmology, Department of Physics and Astronomy, University of California, Irvine, Irvine, CA, USA}
\author[0000-0002-7248-1566]{Z. Cemile Marsan}
	\affiliation{Department of Physics and Astronomy, York University, Toronto, Ontario, Canada}
\author{Ian McConachie}
	\affiliation{Department of Physics and Astronomy, University of California, Riverside, CA 92521, USA}
\author[0000-0001-6251-3125]{Jeffrey C. C. Chan}
	\affiliation{Department of Physics and Astronomy, University of California, Riverside, CA 92521, USA}
\author{Percy Gomez}
	\affiliation{W. M. Keck Observatory, Kamuela, HI, USA}
\author[0000-0002-0332-177X]{Erin Kado-Fong	}
	\affiliation{Department of Astrophysical Sciences, Princeton University, Princeton, NJ, USA}
\author[0000-0003-1181-6841]{Francesco La Barbera}
	\affiliation{INAF--Osservatorio Astronomico di Capodimonte, Napoli, Italy}
\author[0000-0002-2057-5376]{Ivo Labb\'e}
	\affiliation{Centre for Astrophysics \& Supercomputing, Swinburne University of Technology, Hawthorn, VIC, Australia}
\author{Daniel Lange-Vagle}
	\affiliation{Physics Department, Tufts University, Medford, MA, USA}
\author[0000-0002-7356-0629]{Julie Nantais}
	\affiliation{Departamento de Ciencias F\'isicas, Universidad Andr\'es Bello, Santiago, Chile}
\author[0000-0001-6342-9662]{Mario Nonino}
	\affiliation{INAF--Osservatorio Astronomico di Trieste, Trieste, Italy}
\author[0000-0002-0033-5041]{Theodore Pe\~na}	
	\affiliation{Physics Department, Tufts University, Medford, MA, USA}
\author[0000-0003-3959-2595]{Paolo Saracco}	
	\affiliation{INAF--Osservatorio Astronomico di Brera, Milano, Italy}
\author[0000-0001-7768-5309]{Mauro Stefanon}
	\affiliation{Leiden Observatory, Leiden University, Leiden, Netherlands}
\author[0000-0003-1535-2327]{Remco F. J. van der Burg}
	\affiliation{European Southern Observatory, Garching, Germany}

\begin{abstract}

We present spectra of the most massive quiescent galaxy yet discovered at $z>3$, spectroscopically confirmed via the detection of Balmer absorption features in the $H-$ and $K-$bands of Keck/MOSFIRE.
The spectra confirm a galaxy with no significant ongoing star formation, consistent with the lack of rest-frame UV flux and overall photometric spectral energy distribution.
With a stellar mass of $3.1^{+0.1}_{-0.2} \times 10^{11} ~{\rm M}_\odot$ at $z = 3.493$, this galaxy is nearly three times more massive than the highest redshift spectroscopically confirmed absorption-line identified galaxy known.
The star-formation history of this quiescent galaxy implies that it formed $>1000 ~{\rm M}_\odot/$yr for almost 0.5 Gyr beginning at $z\sim7.2$, strongly suggestive that it is the descendant of massive dusty star-forming galaxies at $5<z<7$ recently observed with ALMA.
While galaxies with similarly extreme stellar masses are reproduced in some simulations at early times, such a lack of ongoing star formation is not seen there.
This suggests the need for a more rapid quenching process than is currently prescribed, challenging our current understanding of how ultra-massive galaxies form and evolve in the early Universe.

\end{abstract}

\section{Introduction}

Over the last decade, deeper and wider field near-infrared detected multi-wavelength surveys have enabled the discovery and photometric investigation of rare ultra-massive galaxies (UMGs; $M_*>10^{11} ~{\rm M}_\odot$) at progressively higher redshifts \citep[\textit{e.g.},][]{Rodighiero2007, Wiklind2008, Mancini2009, Marchesini2010, Stefanon2015, Marsan2017}.
Although most UMGs observed at $z>2$ are still forming stars, often quite vigorously \citep{Martis2016, Whitaker2017, TWang2019, Martis2019}, the number of quiescent candidates has been increasing and exceeds the predictions of simulations by a factor of between 3 and 30, depending upon selection criteria \citep[\textit{e.g.},][]{Straatman2014, Guarnieri2019, Pampliega2019}.
A handful of these massive quiescent systems have been spectroscopically confirmed at $1.5<z<2.5$, enabling a more precise characterization of their stellar populations, and improved modeling of their star formation histories due to the detection of stellar continuum features \citep[\textit{e.g.,}][]{Kriek2016, Kado-Fong2017, Belli2019}.

Due to the faintness of such objects at $z>3$, the number of candidates spectroscopically confirmed at these higher redshifts has remained low \citep[][hereafter S18]{Marsan2015, Marsan2017, Glazebrook2017, Schreiber2018a, Schreiber2018b}.
While small, this higher redshift sample suggests that the selection techniques used for these candidates, typically involving rest-frame colors, yield relatively pure samples, though perhaps not complete \citep[][S18]{Marsan2015, Merlin2018}.
The confirmation success rate in S18 also seems to confirm the aforementioned excess relative to simulations is indeed real.

The leading candidates for progenitors of these galaxies, which clearly must form stellar mass at extreme rates at early times, are high-redshift dusty star-forming galaxies (DSFGs).
Recent ALMA observations of small numbers of these DSFGs at $5<z<7$ reveal large amounts of molecular gas and extreme star formation rates \citep[\textit{e.g.},][]{Capak2011, Riechers2013, Riechers2017, Strandet2017, Marrone2018}.
The lack of deep stellar continuum spectra for these $z>3$ UMGs however \citep[3 UMGs with absorption features robustly detected;][S18]{Glazebrook2017} has prevented establishment of a firm link between these objects and the DSFGs, as photometric studies cannot robustly infer the past star-formation history.

In this letter, we present deep rest-frame optical spectra of XMM-2599, a quiescent UMG candidate at $z_{\rm phot}\sim3.4$.
Our spectra confirm its quiescent nature and imply a period of intense star formation ($>1000 ~{\rm M}_\odot / {\rm yr}$) in its $z\sim5.5$ progenitor, consistent with most DSFGs observed at that epoch.
The spectroscopic confirmation of XMM-2599, the most massive quiescent galaxy at $z>3$, arguably represents the biggest challenge yet to the latest theoretical models of galaxy formation in the early Universe, underlining the inadequate quenching mechanism(s) currently implemented in simulations.

Below we describe our target selection and spectral reduction in Section \ref{Data}, our derivation of various galaxy characteristics in Section \ref{Analysis}, and follow with a discussion (Section \ref{Discussion}) and conclusions (Section \ref{Conclusions}).
For this work we assume a Chabrier IMF, $H_0=70 ~\rm{km \textrm{ } s^{-1} Mpc^{-1}}$, $\Omega_m=0.3$ , and $\Omega_\lambda=0.7$.

\section{Data}\label{Data}

\subsection{Target Selection}

\begin{figure*}
\includegraphics[scale=0.5]{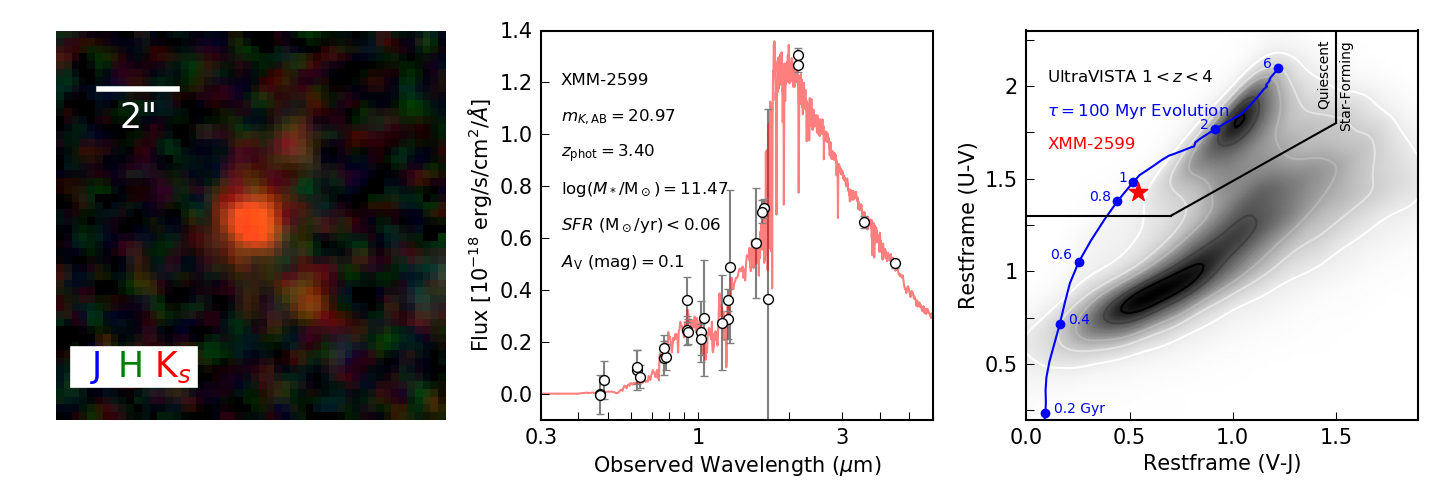} 
\caption{Photometric properties of XMM-2599. \textbf{Left:} Near-infrared imaging of XMM-2599. \textbf{Middle:} Photometric spectral energy distribution of XMM-2599.  Data are shown in white with gray $1\sigma$ errorbars, while the best fit template to the photometry alone is shown in red. Listed properties are also derived from the photometry alone. \textbf{Right:} XMM-2599 on the restframe \UVJ\ diagram. A mass-complete sample of galaxies at $1<z<4$ from UltraVISTA are shown in gray for comparison. The evolution of a population with an exponentially-declining star-formation history parameterized by $\tau=100~\rm{Myr}$ is shown in blue, with several ages labeled in Gyr.  
}
\label{phot}
\end{figure*}

Selected from deep, 28-band imaging catalogs of the VIDEO XMM-Newton field (spanning 0.3-4.5$\mu$m, Annunziatella et al., in prep.), the galaxy XMM-2599 (R.A.$=02^\textrm{h}27^\textrm{m}10.098^\textrm{s}$, Dec.$=-04^\circ 34^{\prime}44.988^{\prime\prime}$) is luminous in the $K_s$-band ($m_{AB} = 20.97^{+0.02}_{-0.02}$), with a narrow singly-peaked redshift probability distribution ($z_{\rm phot} = 3.40^{+0.12}_{-0.10}$), and a spectral energy distribution (SED) consistent with a quenched galaxy (see Figure \ref{phot}, which also lists the stellar population properties derived from SED modeling). 
Taken together, these three characteristics strongly suggest this galaxy is observed when the Universe was only 1.5-2.0 billion years old, has a stellar mass $\log( M_*/{\rm M}_\odot) \sim 11.5$, and is no longer forming stars at an appreciable rate.
As shown in Figure \ref{phot}, the galaxy also lies in the quiescent wedge of the rest-frame \UVc\ vs. \VJc\ (\UVJ) color-color diagram, consistent with the positions of post-starburst galaxies.

\subsection{Spectroscopic Follow-up}

\begin{figure*}
\includegraphics[scale=0.55, trim={0.3cm 0cm 0.3cm 0cm},clip]{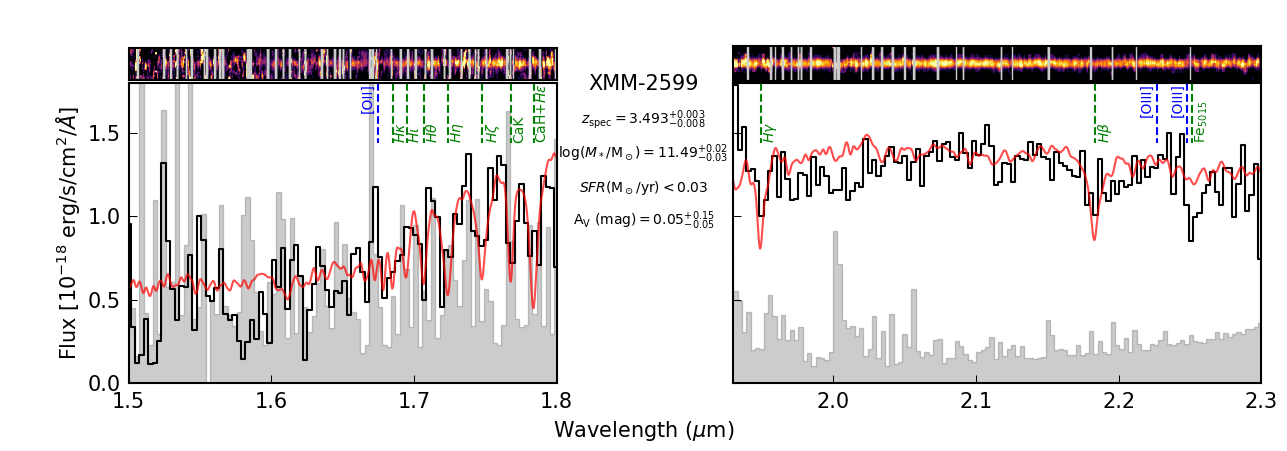}
\caption{Near-infrared H and K band spectra for XMM-2599 and best-fit model.
\textbf{Top:} The telluric corrected 2D spectra, smoothed for visual clarity.  Strong sky lines are masked with gray lines.
\textbf{Bottom:} The 1D extracted spectra, shown in bins $30 ~{\rm \AA}$ wide, are black, while the $1\sigma$ noise (including telluric correction) is gray. The best fit template to the combined photometry and spectroscopy is plotted in red. The location of absorption features are indicated in green, and the wavelengths corresponding to nebular emission from oxygen are blue.
}
\label{spec}
\end{figure*}

We obtained deep spectra of XMM-2599 using the MOSFIRE spectrograph \citep{McLean2010, McLean2012} on the Keck I telescope (PI Wilson; Figure \ref{spec}).
Observations were taken in November and December of 2018.
A single mask was observed in $K$-band for $2^\textrm{h}45^\textrm{m}$, with an average seeing of $0.6^{\prime\prime}$, as determined from a slit star.
Two masks in $H$-band were observed for on-source times of $2^\textrm{h}20^\textrm{m}$ and $2^\textrm{h}40^\textrm{m}$, with seeing of $0.94^{\prime\prime}$ and $1.13^{\prime\prime}$, respectively.

We began reduction by running the MOSFIRE Data Reduction Pipeline\footnote{https://github.com/Mosfire-DataReductionPipeline/MosfireDRP} (DRP) to obtain 2D target and error spectra.
The DRP constructs a pixel flat image, identifies slits, removes thermal contamination ($K$-band), performs wavelength calibration using sky lines, Neon arc lamps, and Argon arc lamps, removes sky background, and rectifies the spectrum.
A custom Python code was written to perform 1D spectral extraction from the DRP outputs utilizing an optimal spectral extraction \citep{Horne1986}.

Additional code was written to perform telluric corrections based on spectra of bright stars ($15<m_{K_s}<18$) included on the MOSFIRE slit masks.
Similar to S18, this code uses the PHOENIX star models \citep{Husser2013} to fit the near-infrared photometry of the stars and thus obtain intrinsic stellar spectra.
The ratio of this model to the extracted 1D spectrum yields a telluric correction which is applied to other objects on the same mask.

The last piece of our reduction entailed identifying and masking out sky lines, which is of critical importance for such faint targets.
To do this, we extracted 1D spectra of the sky from regions of slits that were uncontaminated by any object as determined from inspection of the 2D spectrum and $K$-band imaging.
This resulted in $\sim10$ spectra per mask, which were co-added to create a sky spectrum.
The error spectra for these regions were added in quadrature, excluding wavelength regions of individual spectra that did not fall on the detector.
We then fit and normalized by a `continuum' to the error curve to isolate noise spikes associated with sky lines.
Any pixels on this curve above the 87.5$^\textrm{th}$ percentile of the sky spectra were considered to be strong sky lines, as was any adjacent pixel.
This process reliably identifies sky lines when compared to a visual inspection of a 2D spectrum.
Data from wavelengths affected by sky lines were then masked out for fitting purposes.
For visualizations, sky line pixels were averaged with nearby non-affected pixels to reduce the effects of the sky lines on the spectra, and data were binned.

\subsection{Spectral Features}

The final spectra of XMM-2599 show Balmer series absorption lines redshifted to $z=3.493^{+0.003}_{-0.008}$ (Figure \ref{spec}).
These Balmer lines constrain the age of a galaxy, as they are associated with stars of mass $1.5-2 ~{\rm M}_\odot$, which have main sequence lifetimes of hundreds of millions of years, thereby breaking degeneracies in SED fitting associated with dust and stellar age. 
$\rm{H}\gamma$, $\rm{H}\epsilon + \rm{Ca H}$, $\rm{H}\xi$, $\rm{H}\eta$, and $\rm{H}\theta$ are detected, while $\rm{Ca K}$ lies in a region of significant sky noise.
$\rm{H}\beta$ is seen in absorption, with the possibility of a small emission spike overlaid.
We do not observe nebular emission from oxygen ([O {\rm \scriptsize III}]$\lambda\lambda4959,5007$ and [O {\rm \scriptsize II}]$\lambda\lambda3726,3729$), though the redshifted [O {\rm \scriptsize II}] doublet falls in a region of strong sky emission.

\section{Analysis}\label{Analysis}

\subsection{Galaxy Fitting}

For a consistent comparison with the sample from S18, we utilize the FAST++ code\footnote{https://github.com/cschreib/fastpp} \citep[][S18]{Schreiber2018a} to model the SEDs of our galaxies.
FAST++ is a rewrite of FAST \citep{Kriek2009} for C++ which allows for flexible star-formation history (SFH) parameterizations as well as spectroscopic data of different wavelength resolutions.
Furthermore, spectra are flux scaled to match the observed photometry for individual galaxies, and thus only spectral features / shape contribute to the fit.

The spectrum from each bandpass was fit independently with the photometry to ensure that relative spectral flux calibrations between bandpasses did not affect the outcome.
Both best fit templates were nearly identical, and the spectra were each scaled to match the resultant best fits.
Said scaling differences here were $\sim10\%$.
Finally, each spectrum was allowed to vary relative to the other by up to 2 pixels to account for possible wavelength calibration errors.
We then refit the photometry with the scaled spectra from both bandpasses -- again yielding a best fit template nearly identical to those produced with each band individually.

The grid of potential models tested with FAST++ included those with $3<z<4$, $8.0<\log({\rm age}/\rm{yr})<$ age of the Universe at $z_{\rm model}$, and $0<A_V<5$.
Metallicities of $Z=0.004, 0.008, 0.02,$ and $0.05$ were tested, however the differences in $\chi^2$ between the models of different metallicities are too small to differentiate given the signal-to-noise of our data.
Throughout this work we have quoted results from the $Z=0.02$ (Solar) metallicity run.

\subsection{Star-Formation History}\label{SFH}

Given the ability of FAST++ to fit various functional forms of SFH, we begin with the form presented in S18, which can roughly reproduce the more complex shapes found in best-fit SFHs for massive quiescent galaxies at $z\sim2$ \citep{Belli2019}:
\begin{eqnarray}
SFR_{\rm base}(t) \propto
\begin{cases}
  \rm{e}^{(t_{\rm burst}-t)/\tau_{\rm rise}},& \text{for } t>t_{\rm burst}\\
  \rm{e}^{(t-t_{\rm burst})/\tau_{\rm decl}},& \text{for } t\leq t_{\rm burst}
\end{cases}\\
SFR(t) = SFR_{\rm base}(t) \times
\begin{cases}
  1,& \text{for } t>t_{\rm free}\\
  R_{\rm SFR},& \text{for } t\leq t_{\rm free}
\end{cases}
\end{eqnarray}

This SFH parameterization allows for a period of rising star formation, as well as decoupling the rising and falling exponential phases from the star formation at the time of observation \citep[][S18]{Papovich2011, Glazebrook2017, Schreiber2018a}.
The grid of SFH parameters ranged from $7.0<\log(t_{\rm burst}/\rm{yr})<9.2$, $7.0<\log(\tau_{\rm rise}/\rm{yr})<9.5$, $7<\log(\tau_{\rm decl}/\rm{yr})<9.5$, $7<\log(t_{\rm free}/\rm{yr})<8.5$, and $-2.0<\log(R_{\rm SFR})<5.0$.

\begin{figure*}
\includegraphics[scale=0.75]{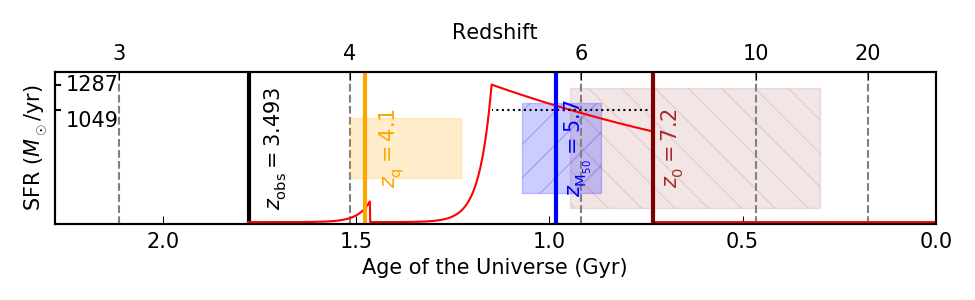}
\caption{Best fit star-formation history for XMM-2599.  The red curve indicates the SFR over cosmic time, with the maximum SFR and a characteristic average SFR shown in solar masses per year on the y-axis.  The black line indicates the spectroscopic redshift and the maroon line is the time that the galaxy began forming stars.
The orange line is the time at which SFR drops below 10\% of the previous average SFR while the blue line denotes the time at which half of the final stellar mass has been formed. Shaded regions correspond to $1\sigma$ confidence intervals.
}
\label{sfh}
\end{figure*}

\begin{figure*}
\includegraphics[scale=0.7]{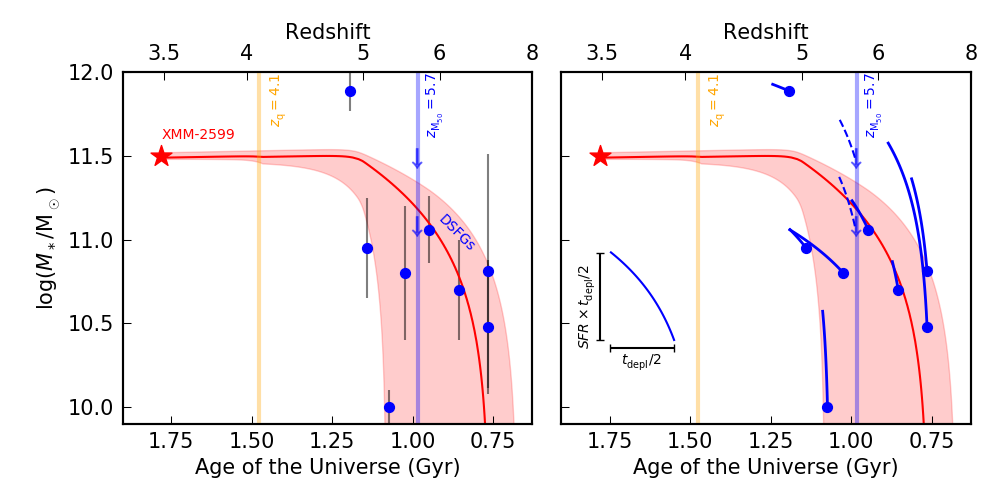}
\caption{High-redshift DSFGs as potential progenitors of XMM-2599.  We show the stellar mass evolution for XMM-2599 in red as calculated from our best-fit SFH, with a shaded 68\% confidence interval. \textbf{Left:} Several high-redshift DSFGs are shown in blue with errors on masses \citep{Capak2011, Riechers2013, Cooray2014, Ma2015, Riechers2017, Strandet2017, Marrone2018, Williams2019, Jin2019}. Reported upper limits are plotted as arrows.  \textbf{Right:} Blue segments show the evolution of the DSFGs assuming the published star formation rate held constant over half the gas depletion timescale (\textit{i.e.}, half of the available gas is turned into stars).  When no gas depletion timescale or gas mass is reported, we set $t_{\rm depl}=0.1$ Gyr, a value typical of the population. The overlap of these tracks with the mass evolution of XMM-2599 suggests that they are potential high-redshift progenitor systems.
}
\label{evo}
\end{figure*}

The best-fit SFH of the form given above implies that this galaxy formed $>1000 ~{\rm M}_\odot/$yr for almost 0.5 Gyr beginning at $z\sim7.2$ (Figure \ref{sfh}).
Our analysis makes use of this SFH, including in the derivation of the mass formation history in Figure \ref{evo}.
However we also fit the data using a variety of other common functional forms of SFH, including exponentially-declining, delayed exponentially-declining, truncated, and top-hat forms.
Aside from the delayed exponentially-declining SFH, which builds stellar mass to unreasonable levels in the early Universe, all functional forms yield similar results, with significant star formation completed by $z\sim5$, and highly suppressed star formation possibly continuing until $z=4-4.5$.

\subsection{Star Formation Rates}

\begin{figure*}
\includegraphics[scale=0.6]{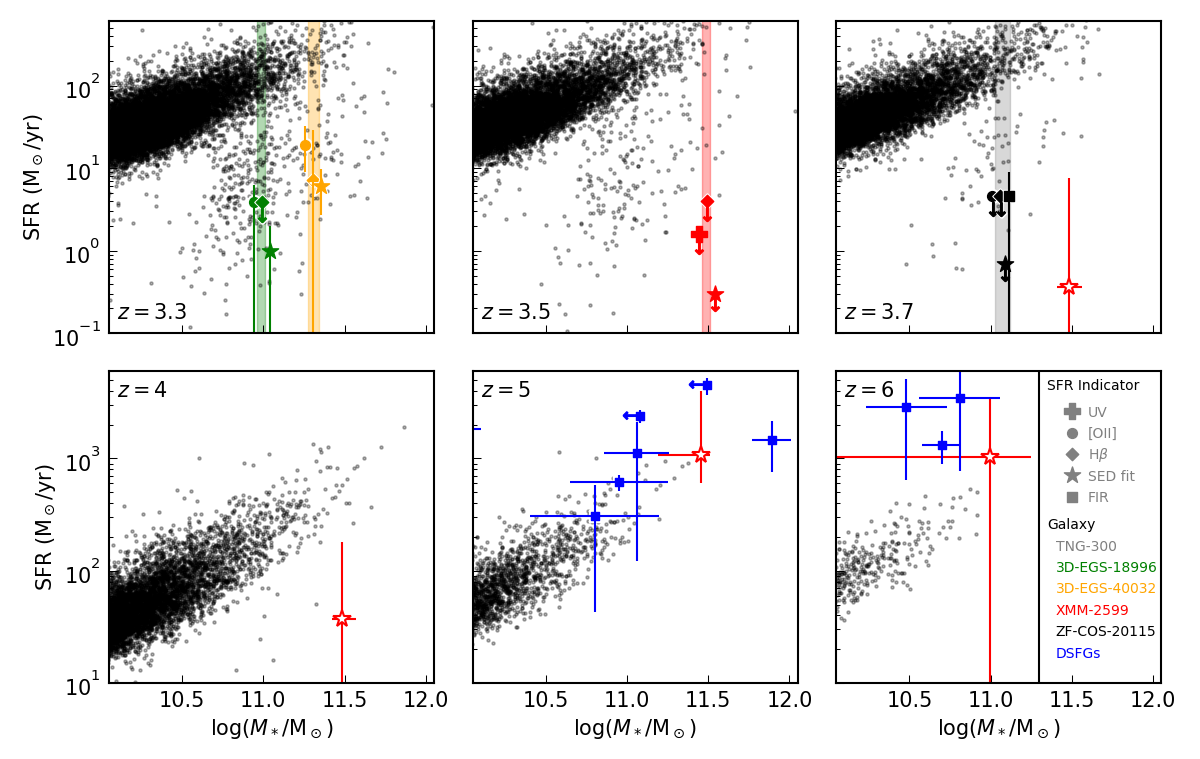}
\caption{Comparison to the Illustris TNG-300 simulation on the SFR-$M_*$ plane.  We show the spectroscopically confirmed absorption-line identified UMGs at $z_{\rm spec}>3$ (green, orange, red, and black), simulated galaxies from six snapshots in Illustris TNG-300 (gray), and the DSFGs from Figure \ref{evo} (blue). Several probes of star formation are shown differentiated by marker style, many as $1\sigma$ upper limits. These are offset along the abscissa for visual clarity, while the best-fit stellar mass is shown as a column, with the width indicating the 68\% statistical error.  Using the best-fit SFH from Figure \ref{sfh}, we plot the position of XMM-2599 at previous epochs as well (open red stars). Note that the range of the ordinate axis differs in the two rows.
}
\label{simcomp}
\end{figure*}

Star formation rates are calculated in several ways for XMM-2599, as shown in Figure \ref{simcomp}.
Values for other UMGs are obtained from S18, and are calculated in the same way.
The SED-derived SFR for XMM-2599 was calculated from FAST++, using the same parameter grid as above.
The ultraviolet SFR is calculated from the best fit SED template by integrating flux density over a $350$\AA\ tophat filter centered on $2800$\AA\ restframe and converting to a star formation rate \citep{Kennicutt1998, Muzzin2013}:
\begin{eqnarray}
SFR_{\rm UV} \textrm{ }[M_\odot \textrm{ yr}^{-1}] = 3.23 \times 10^{-10} L_{2800} \textrm{ }[L_\odot]
\end{eqnarray}
Similar calculations are done to determine SFR based on integrated line fluxes from the MOSFIRE spectra for [O {\rm \scriptsize II}] \citep[][S18]{Kennicutt1998} and $\rm{H}\beta$ \citep[][S18]{Kewley2004}:
\begin{eqnarray}
SFR_{\rm [OII]} \textrm{ }[M_\odot \textrm{ yr}^{-1}] = 1.59\times10^{-8}  L_{\rm [OII]} \textrm{ }[L_\odot]\\
SFR_{\rm H\beta} \textrm{ }[M_\odot \textrm{ yr}^{-1}] = 5.46\times10^{-8}  L_{\rm H\beta} \textrm{ }[L_\odot]
\end{eqnarray}
In the case of XMM-2599, we note that strong emission is not obvious in either case.
$\rm{H}\beta$ may have a small amount of emission overlaid on the stronger absorption feature, while there is strong sky emission on the wavelengths corresponding to [O {\rm \scriptsize II}], yielding a signal-to-noise ratio of $SNR_{\rm{[OII]}}\sim0.2$.
Using the above equations we calculate an upper limit of $SFR_{\rm{H}\beta}<4 ~M_\odot/\rm{yr}$ for XMM-2599, and find that a line flux of $f_ {\rm [OII]}=5.5\times10^{-18} ~\rm{erg/s/cm^2}$ is necessary to reproduce this value.
Assuming an emission feature width of $10$\AA\ and a continuum level of $~{\sim6\times10^{-19} ~\rm{erg/s/cm^2/\AA}}$ from the best-fit SED, this corresponds to a peak line flux density of $f_{\lambda,\rm{[OII]}}~{\sim1\times10^{-18} ~\rm{erg/s/cm^2/\AA}}$.
Although this is broadly consistent with the spectra, we do not plot this value or a limit on Figure \ref{simcomp} due to the very low signal to noise.

\section{Discussion}\label{Discussion}

\subsection{Progenitors of Quiescent UMGs}

In order to build up such a large stellar mass at early times, the progenitors of systems like XMM-2599 must have been explosively star-forming at $z\sim5-6$.
DSFGs at $z>5$ have been confirmed using longer wavelength data, such as that provided by ALMA, but those with large published gas and/or stellar masses remain few \citep{Capak2011, Riechers2013, Cooray2014, Spilker2016, Strandet2017}.
While the low number densities of these DSFGs suggest that they cannot account for all of the quiescent galaxies photometrically identified at $3<z<4$ \citep{Straatman2014}, it seems possible that they could be progenitors of the most massive end of the quiescent UMG population, such as XMM-2599.

In Figure \ref{evo}, we explore this possibility for a sample of high-redshift DSFGs with published stellar masses, molecular gas masses, and star formation rates \citep{Capak2011, Riechers2013, Cooray2014, Ma2015, Riechers2017, Strandet2017, Marrone2018, Williams2019, Jin2019}.
These systems have masses consistent with the mass evolution of XMM-2599 derived from our best-fit SFH.
Additionally, the available gas allows for nearly all of them to reach a stellar mass of \logM$>11$ by $z\sim3.5$ with a plausible star formation efficiency through cosmic time.
While such massive high-redshift DSFGs are rare, their existence implies that other galaxies as massive as XMM-2599 at $z\sim3.5$ exist.
Moreover, though many of these DSFGs have clear optical counterparts, the recent discovery of a significant number of DSFGs at $3<z<8$ with no such counterpart indicates that such galaxies may exist in sufficient numbers to be progenitors of the $z>3$ quiescent UMG population down to even lower masses, and have simply avoided detection thus far \citep{Williams2019, TWang2019}.  

\subsection{Comparison to Simulations}

Quenched galaxies such as XMM-2599 are extremely rare as the stellar mass function for the quiescent population declines steeply at the high-mass end.
Data from the 1.62 deg$^2$ UltraVISTA survey \citep{McCracken2012} implies that quiescent UMGs at $3<z<4$ with \logM$>11$ have a density of $n\sim10^{-5.83} ~\rm{Mpc}^{-3}$, while those with \logM$>11.5$ are estimated to be more than a factor of ten rarer, at $n\sim10^{-6.97} ~\rm{Mpc}^{-3}$\citep{Muzzin2013}.
However, they are observed in numbers significantly higher than those predicted by simulations \citep[see \textit{e.g.}, Figure 14 of][]{Pampliega2019}.
Tens of $z>3$ UMGs have been spectroscopically confirmed via detection of faint emission lines implying ongoing star formation or AGN activity \citep[][S18]{Kubo2015, Marsan2015, Marsan2017}.
However only 3 such systems have robust redshifts from the detection of absorption lines alone: 
ZF-COS-20115 at $z=3.715$ with $M_*=1.15^{+0.16}_{-0.09}\times10^{11} ~{\rm M}_\odot$\citep[][S18]{Glazebrook2017}, 3D-EGS-18996 at $z=3.239$ with $M_*=9.8^{+0.04}_{-0.06}\times10^{10} ~{\rm M}_\odot$ (S18), and 3D-EGS-40032 at $z=3.219$ with $M_*=2.03^{+0.16}_{-0.14}\times10^{11} ~{\rm M}_\odot$ (S18).

Given the low observed number densities, large volume simulations are required for comparison.
In Figure \ref{simcomp} we compare observed absorption line UMGs to simulated galaxies in snapshots from Illustris TNG300 ($302.6 ~\rm{Mpc}$ on a side) \citep{Nelson2018, Pillepich2018, Springel2018, Naiman2018, Marinacci2018}.
TNG300 is able to suppress star formation in massive galaxies at high redshift and easily reproduces 3D-EGS-18996 and 3D-EGS-40032, and reproduces ZF-COS-20115 within the observational errors.
Still, at $z=3.49$ TNG300 has low number densities for high mass galaxies with $\rm{SFR}<5 ~\rm{M_\odot}/\rm{yr}$; $n\sim10^{-6.24} ~\rm{Mpc}^{-3}$ for \logM$>11$ and $n\sim10^{-7.44} ~\rm{Mpc}^{-3}$ for \logM$>11.5$.
Additionally, XMM-2599 remains $\sim1.5-8\sigma$ away from any simulated galaxy of its mass based on the various SFR limit determinations. 

Three possible analogues for ZF-COS-20115 were found in the {\sc{meraxes}} semi-analytic model \citep{Mutch2016a, Qin2017} (box size$=125 ~h^{-1}\textrm{Mpc}$), though none of these approach the mass of XMM-2599.
Other large simulations such as Millenium ($500h^{-1} ~\rm{Mpc}$ on a side) \citep{Springel2005, Henriques2015} do not come close to reproducing any of these quiescent UMGs.
In order to do so simulations require either a more rapid buildup of stellar mass \textit{in situ} during the epoch of reionization or a faster quenching mechanism than is currently prescribed.

We also compare the evolution of XMM-2599 based on our best-fit star-formation history, as shown in Figure 5.
This shows that at $5<z<6$, the characteristics of XMM-2599, \textit{i.e.}, large stellar masses and extreme SFRs, are well reproduced by TNG300.
This is also clear from the ability of TNG300 to reproduce the observed properties of the DSFGs.
However, TNG300 is unable to match the rapidity with which XMM-2599 is quenched at $3.5<z<4$.
Various parametric forms of SFH were tested, as well as different metallicities, and none of these eliminate this issue.

\subsection{Possible Alternatives}

Upon follow-up with high-resolution HST imaging, a number of red, massive, high-redshift galaxies detected with near-infrared ground-based imaging have been revealed to be close pairs \citep{Marsan2019, Mowla2019}.
We lack high-resolution HST imaging for XMM-2599, and thus the case of two compact galaxies in extreme proximity cannot be ruled out.
However we also note that examples of this, as shown in Figures 3 and 4 of \citet{Marsan2019} and Figure 2 of \citet{Mowla2019}, exhibit clear deviations from a compact, circular object in the near-infrared imaging, which XMM-2599 does not (Figure \ref{phot}).

Such close pairs are evidence of future mergers, and therefore XMM-2599 may be the result of a recent dry merger, which lacked sufficient cold gas to trigger substantial star formation.
Future high resolution imaging could pick-up more structural features and shed light on whether this object is the result of a recent dry merger, or indeed a pair of galaxies.
We note that, assuming a 1:1 mass ratio, these galaxies / progenitors would still have stellar masses \logM$\sim11.2$, making them both UMGs.

Nearby neighbors can also contaminate bands with lower spatial resolution, in particular the IRAC bandpasses.
ZF-COS-20115 provides a case study of this, as an optically invisible neighbor led to an initial overestimate of the stellar mass by $\sim 40\%$ \citep{Glazebrook2017, Schreiber2018a}.
While XMM-2599 has two neighbors in the near-infrared ($\sim1.5-2^{\prime\prime}$ away), they are sufficiently distant as to not contaminate the photometry, and the light profile of XMM-2599 is consistent with a roughly circular, singly-peaked distribution perturbed by noise.
Refitting XMM-2599 assuming extreme contamination from these neighbors in IRAC in line with \citet{Schreiber2018a}, \textit{i.e.}, 15\% in $3.6\mu$m imaging and 28\% in $4.5\mu$m imaging, still results in a stellar mass of \logM$\sim11.4$, more massive than any other $z>3$ quiescent UMG.

While massive quiescent populations remain rare at high redshift, star-forming systems in this mass regime, nearly all of which are dust-obscured, are more common \citep{Marchesini2014, Martis2016, Whitaker2017, Martis2019}.
Since heavily dust-obscured galaxies and quiescent galaxies can have similar UV-NIR photometry, it is important to rule out the possibility that XMM-2599 is a dusty galaxy.
Although large amounts of dust can severely dampen emission line signatures in spectra, reproduction of absorption lines by dust is difficult and requires an old stellar population.
Long wavelength data is a certain way to rule out ongoing dust-obscured star formation but the only far infrared imaging in the region, with Herschel-PACS in HerMES \citep{Oliver2012}, shows no detection near XMM-2599.
However, the imaging would only detect objects with $SFR_{IR}>1000 ~{\rm M}_\odot$ and is thus insufficiently deep to constrain the nature of XMM-2599.

ALMA follow-up of massive galaxies at $z>3$ has shown that \UVJ\ color selection also does a good job of identifying truly quiescent galaxies \citep[][S18]{Schreiber2018a}.
XMM-2599 has rest-frame colors \UVc$=1.43^{+0.03}_{-0.02}$ and \VJc$=0.54^{+0.06}_{-0.02}$, thus placing it within the quiescent wedge of the \UVJ\ diagram (Figure \ref{phot}).
More specifically, XMM-2599 lies in the blue corner of the quiescent wedge, typically associated with younger, post-starburst galaxies, as opposed to redder galaxies that quenched in the distant past.
This limits the amount of dust obscuration possible, as substantial dust would move the galaxy toward the red side of the wedge.
Given the lack of emission lines, the superior fit of quiescent galaxy templates to the data, and the rest-frame colors of XMM-2599, all the evidence suggests this galaxy is quiescent.

\section{Conclusions}\label{Conclusions}

In this work we presented spectra confirming the existence of a quiescent galaxy at $z=3.493$ with a stellar mass of $3.1\times10^{11}~{\rm M}_\odot$.
The rest-frame colors combined with the lack of emission lines from nebular oxygen reduce the likelihood of ongoing, dust-obscured star formation.
This galaxy's star-formation history suggests a period of intense star formation, $>1000 ~{\rm M}_\odot/{\rm yr}$ for several hundred Myr at $z\sim6$, consistent with the most gas-rich DSFGs observed at that epoch.

Simulations have improved substantially in the last few years, and are able to reproduce the massive, star-forming DSFGs observed at high redshift that are considered possible progenitors for massive quenched galaxies such as XMM-2599.
However they are still unable to reproduce massive, quiescent galaxies at $z\sim4$.
The specific mechanisms which enable the rapid transformation of these galaxies is unclear, and may in fact be the result of several concurrent events.
While gas-rich major mergers are important in building up the stellar mass at early times, a reduction in the number of these events would limit the amount of gas available for star-formation.
Virial shocks and increased feedback from active galactic nuclei could provide the energy necessary to keep any remaining gas heated, thus prevent the cooling and collapse necessary for forming stars \citep[\textit{e.g.}][]{Man2018}.
Improved ability to replicate these events in the early Universe is required to reproduce this extreme galaxy in simulations.

\section{Acknowledgements}

The authors wish to recognize and acknowledge the very significant cultural role and reverence that the summit of Maunakea has always had within the indigenous Hawaiian community.  We are most fortunate to have the opportunity to conduct observations from this mountain.
This work is supported by the National Science Foundation through grants AST-1517863, AST-1518257, and AST-1815475, by {\it HST} program number GO-15294, and by grant numbers 80NSSC17K0019 and NNX16AN49G issued through the NASA Astrophysics Data Analysis Program (ADAP).
Support for program number GO-15294 was provided by NASA through a grant from the Space Telescope Science Institute, which is operated by the Association of Universities for Research in Astronomy, Incorporated, under NASA contract NAS5-26555.
Further support was provided by the Faculty Research Fund (FRF) of Tufts University and by Universidad Andrés Bello grant number DI-12-19/R.
B.F. thanks L. Alcorn for discussions regarding the MOSFIRE DRP and E. Conant for advice.

\clearpage

\bibliography{/Users/ben/Documents/library}

\end{document}